# UEFI Vulnerability Signature Generation using Static and Symbolic Analyses*

Md Shafiuzzaman[†]
University of California, Santa Barbara
Santa Barbara, CA, USA
mdshafiuzzaman@ucsb.edu

Achintya Desai [†]
University of California, Santa Barbara
Santa Barbara, CA, USA
achintya@ucsb.edu

Laboni Sarker
University of California, Santa Barbara
Santa Barbara, CA, USA
labonisarker@ucsb.edu

Tevfik Bultan
University of California, Santa Barbara
Santa Barbara, CA, USA
bultan@ucsb.edu

## ABSTRACT

Since its major release in 2006, the Unified Extensible Firmware Interface (UEFI) has become the industry standard for interfacing a computer's hardware and operating system, replacing BIOS. UEFI has higher privileged security access to system resources than any other software component, including the system kernel. Hence, identifying and characterizing vulnerabilities in UEFI is extremely important for computer security. However, automated detection and characterization of UEFI vulnerabilities is a challenging problem. Static vulnerability analysis techniques are scalable but lack precision (reporting many false positives), whereas symbolic analysis techniques are precise but are hampered by scalability issues due to path explosion and the cost of constraint solving. In this paper, we introduce a technique called STatic Analysis guided Symbolic Execution (STASE), which integrates both analysis approaches to leverage their strengths and minimize their weaknesses. We begin with a rule-based static vulnerability analysis on LLVM bitcode to identify potential vulnerability targets for symbolic execution. We then focus symbolic execution on each target to achieve precise vulnerability detection and signature generation. STASE relies on the manual specification of reusable vulnerability rules and attacker-controlled inputs. However, it automates the generation of harnesses that guide the symbolic execution process, addressing the usability and scalability of symbolic execution, which typically requires manual harness generation to reduce the state space. We implemented and applied STASE to the implementations of UEFI code base. STASE detects and generates vulnerability signatures for 5 out of 9 recently reported PixieFail vulnerabilities and 13 new vulnerabilities in Tianocore's EDKII codebase.

## KEYWORDS
Static Analysis, Symbolic Execution, Vulnerability Detection, Vulnerability Signatures, Firmware

## 1 INTRODUCTION

UEFI has become an integral component in modern computer systems, replacing the traditional BIOS to provide enhanced support and stability. From 2013 to 2021, over 2.3 billion personal computer systems were shipped with UEFI, showcasing its widespread adoption [43]. Unfortunately, the frequency of firmware attacks targeting UEFI has also increased significantly [36], and there have been many reported examples of such attacks in recent years that affected millions of users [1–9, 11, 12, 35]. These attacks leverage vulnerabilities in UEFI to escalate privileges or gain unauthorized access. Vulnerability exploits in UEFI offer attackers a critical advantage since UEFI operates before the system boots, making exploits resilient to system reboots and operating system cleanups. Consequently, a compromised system remains vulnerable even if the operating system or hard disk is replaced. This inherent resistance to traditional mitigation techniques, combined with UEFI's widespread use, makes it an attractive target for attackers.

Detecting and characterizing vulnerabilities in UEFI is crucial for timely remediation and preventing attackers from generating exploits. However, this task is challenging for several reasons. First, UEFI exploits often arise from silent corruptions that do not crash the system but escalate privileges to perform unauthorized actions [49]. Additionally, the behavior causing the vulnerability may not manifest under normal operation but only under specific or unusual conditions [50]. Furthermore, attackers may exploit a combination of minor weaknesses to compromise the entire system, making detection and characterization even more difficult.

For vulnerability detection and characterization, automated techniques such as fuzzing, symbolic execution, and static analysis have

*This material is based on research sponsored by DARPA under the agreement number N66001-22-2-4037, NSF under grants CCF-2008660, and CCF-1901098. The U.S. Government is authorized to reproduce and distribute reprints for Governmental purposes notwithstanding any copyright notation thereon. The views and conclusions contained herein are those of the authors and should not be interpreted as necessarily representing the official policies or endorsements, either expressed or implied, of the U.S. Government.
[†]Both authors contributed equally to this research.

been used in many domains. However, firmware development, particularly for UEFI, faces unique challenges due to its specific execution environment and the extensive interaction with hardware [48]. Existing approaches exhibit trade-offs that hinder their efficacy in comprehensive firmware analysis. Static analysis techniques, while scalable, often lack precision and can produce false positives due to their approximation methods [21]. On the other hand, symbolic execution offers precision but struggles with scalability and harness generation when applied to the complex firmware codebase with numerous hardware interactions [28]. The architecture of UEFI also presents challenges to traditional fuzzing techniques, as most UEFI vulnerabilities manifest as silent corruptions rather than crashes, making them difficult for fuzzers to detect [49]. Furthermore, existing fuzzing tools do not account for the interdependencies between different UEFI entry points. They also lack understanding regarding the complex input structures and execution contexts specific to the targets they are testing [49].

In this paper, we propose a scalable and precise technique called STatic Analysis guided Symbolic Execution (STASE) for UEFI vulnerability detection and signature generation. Figure 1 shows the components of the STASE framework. The core of our approach is to employ rule-based static analysis to guide symbolic execution. STASE starts with detecting potential vulnerabilities using rule-based static analysis and then generates vulnerability signatures using symbolic execution. The static analysis rules are derived from common vulnerability patterns and are extensible by the developers' domain knowledge. For every potential vulnerability detected by static analysis, STASE outputs a vulnerability description containing the entry points, attacker-controlled sources, vulnerability locations, affected program instructions (sink), and an assertion for confirming the vulnerability. This vulnerability description is then used to generate the symbolic execution harness, which eliminates the manual intervention required to run symbolic execution and significantly improves its scalability by directing it towards specific potentially vulnerable locations in the program. Finally, symbolic execution generates the vulnerability signatures composed of the preconditions and postconditions of the vulnerabilities. Our overall technical contributions can be summarized as follows:

(1) **STatic Analysis guided Symbolic Execution (STASE)**: A novel approach for combining static analysis that reduces the false positives generated by static analysis and improves the scalability of symbolic execution.
(2) **Rule-based static analysis for UEFI**: A rule-based and extensible static analysis approach to detect potential UEFI vulnerabilities.
(3) **Automated symbolic execution harness generation via rule-based static analysis**: An automated technique for generating symbolic execution harnesses in order to deploy symbolic execution without manual intervention.
(4) **Scalable vulnerability signature generation for UEFI using symbolic execution**: Guiding symbolic execution engine using static analysis generated vulnerability descriptions and harnesses, which effectively narrows the execution path, improving scalability and reducing analysis time, and outputs the vulnerability signatures that identify pre and post-conditions of detected vulnerabilities.

## 2 UEFI VULNERABILITY SIGNATURES

UEFI comprises millions of lines of code from multiple sources, starting with EDKII [44], the open-source reference UEFI implementation from TianoCore, motherboard firmware companies, and motherboard manufacturers that often include unnecessary bloat due to its complex supply chain [24]. This bloat significantly increases the vulnerability surface of UEFI. These vulnerabilities pose significant security threats due to their position in the system boot process and their capability to override security protections. UEFI vulnerabilities stem from various sources, such as buffer overflows, improper handling of UEFI variables, and flaws in System Management Mode (SMM) drivers. Such vulnerabilities can be leveraged to execute unauthorized code, bypass security protocols, or compromise the system's integrity [36].

The SMM drivers are designed exclusively for UEFI and are not intended for application software or general-purpose systems software. The primary advantage of SMM is its distinct and isolated processor environment, which is invisible to the operating system and applications. This isolation is achieved by using System Management Interrupts (SMI) as the only entry point to enter SMM. In SMM, code is executed in a dedicated address space known as SMRAM, which must be kept inaccessible to other CPU modes by the firmware. However, vulnerabilities within SMM drivers can invalidate this protection, leading to real-world attacks [34, 35, 37, 38, 47].

Alongside these SMRAM vulnerabilities, seemingly innocuous buffer overflows or integer underflows can escalate into significant vulnerabilities in UEFI. A notable instance is the set of vulnerabilities known as PixieFail, which was recently disclosed by Quarkslab [39]. These vulnerabilities are found within the network module of EDKII. PixieFail encompasses issues like buffer overflows, out-of-bounds reads, and integer underflow, affecting functionalities during the Preboot eXecution Environment (PXE) stage. These flaws can lead to remote code execution, denial-of-service attacks, DNS cache poisoning, network session hijacking, and data theft [39].

As an example, let us discuss an "SMRAM Write" UEFI vulnerability. This type of vulnerability arises from improper handling of memory within the SMM, where non-SMM code, such as the operating system kernel or user applications, is able to write to SMRAM. To identify such vulnerabilities, we can define conditions (stated as assertions) that characterize the intended behavior of the system, such that their violation indicates a vulnerability. Negation of such an assertion can be used to identify the "vulnerability condition" which indicates a vulnerability when it holds. The vulnerability condition for the SMRAM Write can be expressed as follows: $\neg(\texttt{BufferSize} \leq \texttt{SMRAM\_BASE} + \texttt{SMRAM\_SIZE} \land \texttt{Buffer} \leq \texttt{SMRAM\_BASE} + \texttt{SMRAM\_SIZE} \land (\texttt{BufferSize} = 0 \lor \texttt{Buffer} + \texttt{BufferSize} \leq \texttt{SMRAM\_BASE} + \texttt{SMRAM\_SIZE}))$
This condition validates whether or not the buffer provided to the SMI handlers overlaps with SMRAM. The negation here reflects the violation of the expected condition.

Consider the code snippet in Listing 1 from `SmramProfileRecord.c` file of EDKII. This code performs a write operation on the buffer (`CommBuffer`) passed to an SMI handler (`SmramProfileHandler`). Writing to the `CommBuffer` through the `SmramProfileHandler` function without any overlap check could potentially overwrite SMRAM if the `CommBuffer` overlaps with SMRAM, leading to unintended behavior.

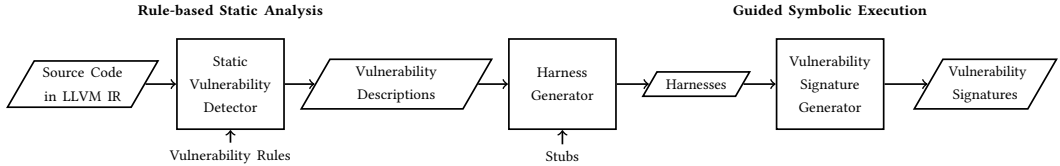

Figure 1: STASE Technique

We can characterize this unintended behavior as a vulnerability signature that captures the preconditions required to reach the vulnerable statement, the code segment containing the vulnerability, and the postconditions resulting from the vulnerability.

```
2165  SmramProfileHandlerGetInfo (
2166    IN SMRAM_PROFILE_PARAMETER_GET_PROFILE_INFO  *
            SmramProfileParameterGetInfo) {  ...
2181    SmramProfileParameterGetInfo->ProfileSize       =
            SmramProfileGetDataSize ();} ...
2290  /*@param CommBuffer A pointer to a collection of data in memory
            that will be conveyed from a non-SMM environment into an
            SMM environment.
2291    @param CommBufferSize   The size of the CommBuffer.*/
2300    ...
2301  SmramProfileHandler ( IN EFI_HANDLE DispatchHandle, IN CONST
            VOID *Context OPTIONAL,IN OUT VOID *CommBuffer OPTIONAL,
            IN OUT UINTN *CommBufferSize OPTIONAL) {  ...
2341      switch (SmramProfileParameterHeader->Command) {
2342      case SMRAM_PROFILE_COMMAND_GET_PROFILE_INFO:   ...
2349      SmramProfileHandlerGetInfo ((
            SMRAM_PROFILE_PARAMETER_GET_PROFILE_INFO *)(UINTN)
            CommBuffer);
2350        break;
2351        ...}
```

**Listing 1: Example of SMRAM Write Vulnerability**

Assume that we added an SMRAM Write vulnerability condition on Line of Code (LOC) 2181 and received an assertion failure.

**Precondition:** The precondition defines the necessary conditions that must be satisfied to navigate from the entry point of the SMM driver (LOC 2301) to the vulnerability location (LOC 2181), where an assertion violation occurs. The path to this vulnerability is characterized by a series of constraints on the system state. An example path constraint is `CommBuffer->Header.Command = 1` and `*CommBufferSize == 24` and `mSmramReadyToLock == 0`

**Code Segment:** This code segment is accessing and manipulating the `CommBuffer` that starts from the entry point at LOC 2301 in `SmramProfileRecord.c` and proceeds to the assertion check at LOC 2181.

**Postcondition:** The postcondition describes the system state when the assertion fails at LOC 2181, indicating a failure to maintain the integrity constraints regarding SMRAM boundaries.

**SMRAM Write:** ¬(*CommBufferSize ≤ SMRAM_BASE + SMRAM_SIZE ∧
CommBuffer ≤ SMRAM_BASE + SMRAM_SIZE ∧ (*CommBufferSize = 0
∨ CommBuffer + *CommBufferSize   ≤ SMRAM_BASE + SMRAM_SIZE))

**Vulnerability Signature:** Putting it all together, the vulnerability signature for this vulnerability can be expressed as a Hoare triple:

$$\{Precondition\} \; CodeSegment \; \{Postcondition\}$$

## 3 RULE-BASED STATIC ANALYSIS

Application of traditional static analysis techniques to vulnerability detection results in many false positives due to their reliance on sound approximation methods [21]. In recent years, rule-based static analysis techniques have been successful in increasing the precision of vulnerability analysis, while enabling their specialization to different domains. Rule-based static analysis approaches rely on domain specific languages for rule specification. Datalog, which is a declarative logic programming language that is a syntactical subset of Prolog, has been widely used for specification of program analysis rules [14, 16, 30, 41, 45]. Furthermore, program analysis queries correspond to fixed-point computations which can be captured as evaluations of logic programs. Hence, once static analysis rules are expressed in Datalog, a Datalog engine can be used to compute the results of the static analysis. In this paper, we use Soufflé [33, 42], which is an expressive dialect of Datalog and provides a performant logic program evaluation engine for efficient rule-based static analysis. Furthermore, we use open-sourced cclyz-erpp [32] tool built on Soufflé which specifies rules for field and structure-sensitive pointer analysis for LLVM-IR source code.

### 3.1 Semantics and Rules

In this section we briefly overview the LLVM-IR semantics and describe Datalog-based static analysis rules.

*3.1.1 LLVM-IR Semantics.* We define the stack frames and program states based on the formalization of LLVM-IR [51] as follows:

*Definition 3.1.* A frame $\Sigma$ is defined as $(f, l, c, t, \Delta, \alpha)$ where $f$ tracks the current function corresponding to the frame, $l$ is LLVM block label, $c$ is the next instruction in the current LLVM block to be executed next, $t$ is termination instruction in the current block, $\Delta$ is the mapping function to keep track of the local variables to their values in LLVM and $\alpha$ is memory block stack that was allocated by 'alloca' instructions within the function.

*Definition 3.2.* A program state $S_P \in \mathbb{S}_P$, where $\mathbb{S}_P$ is the state space for a given program $P$, is defined as $S_P \equiv (M, \overline{\Sigma})$ where $M$ is a memory state as defined in [51] and $\overline{\Sigma} = \{\Sigma_1, \ldots, \Sigma_{nc}\}$ is a stack of frames, with $\Sigma_1$ being top, that keeps track of local variables and current instruction, where $nc$ is the number of frames in the stack.

Let $\text{Attr}(S_P)$ and $\text{Val}(S_P)$ be two sets defined over a program state $S_P$ that denote all the attributes and their values in $S_P$. For example, the function field of the current frame $\Sigma.f$ is a member of $\text{Attr}(S_P)$ and its value, say main, is a member of $\text{Val}(S_P)$.

We define facts and relations based on the program states:

*Definition 3.3.* A relation $Q_T$ is defined as the collection of attributes $\{a_1, \ldots, a_n\}$ representing a program property $T$ with each attribute being individually mapped to an attribute of the program state: $Q_T = \{a_1, \ldots, a_n\}$ where $\forall i \in [n], a_i \in \text{Attr}(S_P)$.

*Definition 3.4.* A fact $\text{F}(Q_T)$ is defined as a set of tuples that correspond to the values of the attributes in relation $Q_T$ in some program state:

$$\text{F}(Q_T) = \left\{(b_1, \ldots, b_n) \mid \exists S_P \in \mathbb{S}_P, \forall i \in [n], a_i \leftarrow b_i \land b_i \in \text{Val}(S_P) \land a_i \in Q_T\right\}$$

For example, the membership property of a function with respect to an LLVM instruction can be expressed as relation $\{a_1, a_2\}$ where $a_1 = S_P \cdot \Sigma_1 \cdot c$ and $a_2 = S_P \cdot \Sigma_1 \cdot f$ where $S_P \in \mathbb{S}_P$ is a program state. Fact, $F(\{S_P \cdot \Sigma_1 \cdot c, S_P \cdot \Sigma_1 \cdot f\})$ say instr_func, would be the collection of values for the instruction and function attributes in program states. A fact can be considered as the set of witnesses for the attribute relation.

```
1  ; Function Attrs: noimplicitfloat noinline noredzone nounwind
       uwtable
2  define dso_local %struct.MEMORY_PROFILE_CONTEXT_DATA*
       @GetSmramProfileContext() #0 {
3    %1 = load %struct.MEMORY_PROFILE_CONTEXT_DATA*, %struct.
       MEMORY_PROFILE_CONTEXT_DATA** @mSmramProfileContextPtr,
       align 8
4    ret %struct.MEMORY_PROFILE_CONTEXT_DATA* %1
5  }
```

**Listing 2: An example function in LLVM-IR**

Consider the LLVM-IR code snippet from Listing 2. Based on the code snippet, instr_func fact will hold the following fact:

(<SmramProfileRecord.bc>:GetSmramProfileContext:0,
<SmramProfileRecord.bc>:GetSmramProfileContext)

where the first entry represents LLVM instruction (index 0) on line 3, and the second entry represents the function to which the instruction belongs.

*3.1.2 Rules for Static Analysis.* In Datalog-based static analysis, rules are specified as horn clauses. Formally, a declarative static analysis rule is defined as follows (see [42] for rule syntax):

*Definition 3.5.* We define a rule $R(C)$ as a horn clause over a set of terms $C = \{c_1, \ldots, c_n\}$ where term $c_1 = \{q_1, \ldots, q_n\}$ is a collection of attributes used to represent the underlying program property, and the rest are either a relation or negation of a relation or a constraint or a rule in itself: $R(C) \equiv c_1 \leftarrow c_2 \land c_3 \ldots \land c_n$ where rule $R(C)$ means that: If $c_2 \land c_3 \ldots \land c_n$ holds, then $c_1$ can be inferred.

For a given rule $R(C)$, the set of program states where $c_1$ can be inferred is categorized as rule qualifying set $P_{\text{set}}^{R(C)}$ under rule-relations. Internally, the datalog engine views these rules as relations and computes them over the facts and constraints till a fixed point is reached which results in $P_{\text{set}}^{R(C)}$ for the corresponding rule. The attributes of these states are extracted and mapped onto the vulnerability descriptions defined at the end of this section.

## 3.2 Vulnerability Rules for UEFI

Our rule-based static analysis for vulnerability detection involves specifying vulnerability rules, taint tracking, and program slicing in order to generate vulnerability descriptions which are used for symbolic execution harness generation.

*3.2.1 Vulnerability Rules.* To perform vulnerability analysis using rule-based static analysis, we first need to specify vulnerability rules. We use the facts from cclyzerpp [17, 32] on LLVM-IR code to design our own vulnerability rules for UEFI that are customizable and reusable. In addition to writing rules for known vulnerability types such as division by zero, buffer overflow, we also add UEFI-specific rules to detect specification violations that are known to lead to vulnerabilities from CVEs related to UEFI. As shown in the work [18], there are more than ten categories of vulnerabilities present in UEFI, with each category having its variations and points of exploit. For STASE, we added six memory related vulnerabilities and three UEFI-specific vulnerabilities, as listed here:

- SMRAM Read vulnerability (based on CVE-2022-35896)
- SMRAM Write vulnerability (based on CVE-2022-23930)
- SMM Callout (based on CVE-2022-36338)
- Integer Underflow vulnerability (based on PixieFail CVE-2023-45229)
- Integer Overflow vulnerability
- Division by zero vulnerability
- Buffer Overflow vulnerability
- Out of bounds access vulnerability
- Use after Free vulnerability

Our vulnerability rules are expressed over facts and constraints. Facts can be seen as basic building blocks for our vulnerability rules, enabling us to perform static vulnerability analysis. To generate facts, we use cclyzerpp's built-in LLVM passes that populate fact relations with program facts based on the abstract syntax tree (AST) of the LLVM modules. From our previous example, instr_func(?instr, ?func) is a fact relation named instr_func which holds the facts as an ordered tuple (?instr, ?func) where ?instr and ?func are arguments that represents an instruction in LLVM and the corresponding function to which the instruction belongs. These fact relations provide a way to abstract out program-specific notions such as variable name, function name, etc., from vulnerability rules. This plays an important role in making our vulnerability rules generalizable and customizable.

In isolation, facts are insufficient to perform complex vulnerability analysis. As defined in the previous section, a vulnerability rule allows us to add conditions that facts must satisfy to qualify as a program property. These conditions are written in terms of fact relations, other vulnerability rules, or constraints over attributes. For example, we capture the three LLVM instruction variations for division by defining a rule divisioninstructions(?divid, ?divis, ?instr) where ?instr is an argument that specifies a potential division instruction in the program, ?divid and ?divis are variables that capture the dividend and divisor associated with the instruction ?instr respectively. We define the rule to capture udiv instruction as follows:

divisioninstructions(?divid, ?divis, ?instr):-
  udiv_instr(?instr),
  udiv_instr_first_operand(?instr,?divid),
  udiv_instr_second_operand(?instr,?divis).

where udiv_instr(?instr) is a fact relation that contains udiv instructions within the LLVM file. udiv_instr_first_operand(?instr, ?divid) and udiv_instr_second_operand(?instr,?divis) are also fact relations that match the udiv instruction to the first and second operator respectively. Based on LLVM syntax, the first and second operands of udiv instruction map to dividend and divisor, respectively. We follow the same process to extend the above rule to include sdiv and fdiv instructions. The divisioninstructions rule can also be extended to accommodate any function calls to functions that are known to perform division internally on the arguments using substr constraint from Soufflé.

Based on this rule, we define our divide by zero vulnerability rule by adding more facts such as instr_pos that maps the LLVM

instruction to the line number in the source code using LLVM debug information. The vulnerability rule is as follows:

```
division_primitive(?func, ?divid, ?divis, ?instr, ?line):-
  instr_func(?instr, ?func),
  divisioninstructions(?divid, ?divis, ?instr),
  instr_pos(?instr,?line,?col).
```

Notice that we do not perform any value flow analysis to detect whether the divisor can potentially be zero or not. This choice is deliberate to keep the tool lightweight and scalable; however, it is likely to result in many false positives, which are filtered out during the symbolic execution phase. It is important to emphasize that the vulnerability rules remain the same as the facts change from one program to another. Hence, once a vulnerability rule is written, it can be reused for analyzing all UEFI code going forward.

*3.2.2 Attacker-controlled Taint Tracking.* One missing piece in the vulnerability rules discussed above is the vulnerability's exploitability. It might be possible to find a precondition that triggers a division by zero vulnerability (i.e., sets the divisor to 0), but an attacked may not be able to trigger it by providing inputs that satisfy the precondition (i.e., attacker may not able to control the value of the divisor). For detection of exploitable vulnerabilities, we can disregard cases that cannot be triggered by an attacker. To make our vulnerability analysis focus on exploitable vulnerabilities, we add attacker-controlled taint-tracking to our vulnerability rules.

To perform taint tracking, we define attacker entry points and attacker-controlled inputs. Attacker entry points are the source code functions that the attacker is able to call. Attacker-controlled inputs are variables in the source code that the attacker can control when it reaches the entry point. Usually, the attacker-controlled inputs are the arguments of the entry point functions or global variables. For the SMI handler domain, triggering a specific SMI handler through the kernel module can be considered an entry point, usually done by writing a specific value to a port designated to trigger SMIs.

In our vulnerability rules, we make SMI handlers as our attacker entry points. We also know that SMM code is kept protected in the SMRAM region, and the only way to interact with it is through a communications buffer, which are the `CommBuffer` and `CommBufferSize` parameters of the SMI handler. Hence, we make `CommBuffer` and `CommBufferSize` as our attacker-controlled inputs for an SMI handler. Now, we can tune our vulnerability rules to capture potential vulnerabilities triggered through the contents of `CommBuffer` and `CommbufferSize` from the SMI handler. The attacker entry points and attacker-controlled inputs are manually defined based on domain specifications and CVE descriptions.

For example, we add SMI handler named `SmramProfileHandler` as an attacker entry point with `CommBuffer` as attacker-controlled input by adding the following rules:

```
entrypoint(?func):-
  func_name(?func, "@SmramProfileHandler").
entryinput(?taintentry):-
  entry_point(?func),
  func_param(?func, ?taintentry, 2).
```

where `func_param` is a fact relation that maps a function to its argument at the given index.

For taint-tracking, we use attacker-controlled input as the taint source. We define taint-sink as a variable at the vulnerable instruction that enables the attacker to influence the said vulnerability. Taint sink differs from one vulnerability rule to another due to the varying vulnerability behaviors. For example, the taint sink for a division by zero vulnerability is its divisor operand, whereas the taint sink for an integer overflow vulnerability could be either of the operands.

To perform taint-tracking, we use pointer analysis of cclyzerpp, which is based on field and structure-sensitive Andersen-style pointer analysis [16]. Using the points-to information, the taint-tracking rule adds the condition that both taint-source and taint-sink should point to the same allocation site in the program. The taint-tracking rule is written as follows:

```
tainttacking(?taintsource,?taintsink):-
  entryinput(?taintsource),
  subset.var_points_to(_,?samealloc,_,?taintsource),
  subset.var_points_to(_,?samealloc,_,?taintsink).
```

Then the modified division-by-zero vulnerability rule with taint-tracking is specified as follows:

```
divisor_tainted_division_primitive(?func, ?divid, ?divis, ?taintsource, ?instr, ?line) :-
  tainttacking(?taintsource, ?divis),
  instr_func(?instr, ?func),
  divisioninstructions (?divid, ?divis, ?instr),
  instr_pos(?instr,?line,?col).
```

Notice that the flow-insensitive nature of pointer analysis from cclyzerpp results in over-approximation. It is possible that the taint-source and taint-sink no longer point to the same allocation site at the potentially vulnerable location. However, taint-tracking still reduces the number of false positives and contributes to the vulnerability description used for symbolic execution harness generation for each potential vulnerability by providing a starting point for symbolic execution in the form of an entry point, potential symbolic variables in the form of taint-sources, and variables to be used in assertion template in the form of taint-sinks.

*3.2.3 Program Slicing.* To run symbolic execution for each potential vulnerability, we want to identify the program locations that can be safely stubbed out. Since symbolic execution faces path explosion, it is crucial that we identify and remove program behaviors that are not related to the vulnerability target. We use program slicing [46] to achieve this. The slicing criteria are determined by the potentially vulnerable instruction and the taint-sink. We implemented the two-pass program slicing algorithm from [31], which is based on the dependence graph. We construct the dependence graph using control dependencies from [26] and data dependencies by tracking def-use chains and memory read-writes using pointer analysis from cclyzerpp. Note that our slicing implementation currently does not support mutual recursion, which can be added using the approach discussed in [15].

*3.2.4 Vulnerability Descriptions.* Once the static vulnerability detector identifies a potential vulnerability target, it produces a vulnerability description. For each vulnerability target, a vulnerability description provides target-specific information as defined below. Later, the harness generator component uses the vulnerability description to guide symbolic execution for each vulnerability target.

*Definition 3.6.* Given a program $P$ and a series of program states $(S_P^e, \ldots, S_P^t)$ where $S_P^e$ is the program state at the attacker-controlled entry point of program $P$ and $S_P^t$ is the program state at vulnerability target, we define the vulnerability description as a 7-tuple representation $V_{S_P^t} = \langle P, E, I, A, K, L, U \rangle$ where

- $P$ is the program in which the potential vulnerability exists
- $E$ is the attacker-controlled entry point, which is defined as $E = S_P^e \cdot \Sigma_1 \cdot f$
- $I$ is the attacker-controlled inputs, which are defined as $I = \{v_1, \ldots, v_m \mid \forall i \in [m],\ v_i = S_P^e \cdot \Sigma_1 \cdot \Delta \cdot v_i\}$
- $A$ is the assertion template corresponding the vulnerability category, which is defined as $A = \text{assert}(\text{condition}(l_1, \ldots l_n))$ such that $\forall i \in [n], l_i = S_P^t \cdot \Sigma_1 \cdot \Delta \cdot v_i$
- $K$ is the vulnerable LLVM instruction, which is defined as $K = S_P^t \cdot \Sigma_1 \cdot c$
- $L$ is the vulnerability location in the source code, which is defined as $L = S_P^t \cdot \Sigma_1 \cdot c \cdot line$
- $U$ is the list of source code locations that can be safely stubbed out, which is defined as $U = \{S_P^i \cdot \Sigma_1 \cdot c \cdot line \mid S_P^i \in (S_P^e, \ldots, S_P^t) \wedge S_P^i \notin \text{Slice}(S_P^e, \ldots, S_P^t)\}$

Based on the code snippet in listing 3, the vulnerability description generated for the target is as follows: $\langle P, E, I, A, K, L, U \rangle$

$P = \text{injected\_Tcg2Smm.c}, E = \text{TpmNvsCommunciate},$
$K = <\text{injected\_Tcg2Smm.bc}>: \text{TpmNvsCommunciate}: 32,$
$A = \text{assert}(\text{TempCommBufferSize} \neq 0), I = \text{CommBufferSize},$
$L = \text{injected\_Tcg2Smm.c}: 70, U = \{\text{injected\_Tcg2Smm.c}: 60\}$

```
47  EFI_STATUS EFIAPI TpmNvsCommunciate (IN EFI_HANDLE
         DispatchHandle, IN CONST VOID  *RegisterContext, IN OUT
         VOID *CommBuffer, IN OUT UINTN *CommBufferSize){
68    ...
69    TempCommBufferSize = *CommBufferSize;
70    mMcSoftwareSmi = mMcSoftwareSmi/TempCommBufferSize;
71    ...
72  }
```

**Listing 3: Injected Division by Zero Vulnerability Example**

## 4 HARNESS GENERATION

The symbolic execution harness establishes the environment and supplies the necessary context for symbolic execution. It has two main components: the environment configuration and the path exploration guidance. Environment configuration involves setting up the system's initial state and context to reflect realistic execution conditions. This includes modeling the external libraries, system calls, and hardware interactions. Path exploration guidance involves directing the symbolic execution toward the program behaviors likely to expose bugs or vulnerabilities. This includes identifying suitable entry points and exit conditions, determining symbolic and concrete variables, and adding bounds on depth and loops.

A STASE symbolic execution harness consists of an Environment Configuration Harness (ECH), which is created once for UEFI vulnerability analysis, and a Path Exploration Harness (PEH), which is automatically generated using the vulnerability descriptions generated during the static analysis phase of STASE. Typically, developers generate these harnesses manually which requires in-depth knowledge of symbolic execution and the system under investigation. STASE automates this process, improving the usability and scalability of symbolic execution.

### 4.1 Environment Configuration Harness (ECH)

STASE leverages basic UEFI domain knowledge, which can be extracted from UEFI specification [29], and provides an Environment Configuration Harness (ECH) for UEFI vulnerability analysis. This sets up the symbolic execution configuration once, establishing a baseline environment that accurately emulates UEFI configurations and hardware interactions.

*4.1.1 Global Table:* Global system and services tables (e.g., EFI System Table, Boot Services Table, and Runtime Services Table) provide standardized interfaces for UEFI applications, drivers, and the underlying system to interact. However, direct interaction with these global tables is neither practical nor necessary during symbolic analysis because the symbolic execution environment does not execute the entire UEFI runtime environment at once. Instead, communication and data manipulation are abstracted, focusing on the program logic. Internal functions or variables typically accessed via global tables are exposed through file references. This abstraction is essential for accurately simulating the firmware's behavior in a controlled environment, allowing the analysis to concentrate on the code's logic and functionality without being constrained by the specifics of the UEFI execution environment.

*4.1.2 PCD-dependent Variable:* The Platform Configuration Database (PCD) enables developers to customize the firmware for specific operational requirements. However, during the symbolic analysis, the exact values of PCD-dependent variables may be unknown. Moreover, relying on a particular PCD value for symbolic execution can limit the analysis to predetermined configurations, potentially missing vulnerabilities that may arise under different settings. To address these limitations and enhance the comprehensiveness of the symbolic execution, symbolic values are assigned to PCD-dependent variables. This enables PCD-dependent variables to simulate various configuration scenarios the firmware might encounter in real-world settings. For instance, the PCD variable `PcdReclaimVariableSpaceAtEndOfDxe` determines whether a certain memory space is reclaimed at the end of the DXE phase. By assigning a symbolic value to this variable, the symbolic execution can explore both scenarios: one where the memory space is reclaimed and one where it is not.

*4.1.3 Firmware Parameter:* Assigning symbolic values to firmware parameters decouples the analysis environment from the hardware infrastructure. This approach allows the symbolic execution engine to explore a broader range of hardware configurations without relying on a fixed setup. For example, the SMRAM base address and size, which dictate the memory region reserved for system management functions, can be treated as symbolic. By doing so, the analysis can consider various possible values for the SMRAM base address and size, enabling the exploration of different memory configurations. Similarly, other critical parameters on memory boundaries, such as `mSmmMemLibInternalMaximumSupportAddress`, can also be treated symbolically to explore their implications on firmware security. This approach ensures comprehensive examination of the firmware under diverse conditions, identifying potential vulnerabilities that may not be apparent with static configurations.

*4.1.4 Protocol GUID:.* A GUID (Globally Unique Identifier) in UEFI is a 128-bit number that is used to identify items such as devices and protocols. Protocol GUIDs are essential for identifying and interacting with the protocols defined within the UEFI specification. Since symbolic execution harness abstracts parts of the system behavior, during symbolic execution specific runtime values for the protocol GUIDs are not generated. Predefining these GUIDs with default values ensures that the symbolic execution engine can accurately simulate the firmware's interactions with various UEFI protocols. The UEFI Specification documents [29], maintained by the UEFI Forum, contain detailed information about protocol GUIDs. These specifications include a list of protocol GUIDs along with their corresponding values and descriptions. STASE UEFI ECH contains the default values for protocol GUIDs obtained directly from the inf files (Information Files that are used to describe the metadata and build configuration for UEFI modules or drivers).

## 4.2 Path Exploration Harness (PEH)

STASE uses static analysis output to automate the path exploration harnesses. As discussed earlier, each potential vulnerability detected by the static analysis phase of STASE is characterized as a vulnerability description $V_{S_P^t} = \langle P, E, I, A, K, L, U \rangle$. Vulnerability descriptions are mapped to the symbolic execution harness to guide symbolic execution towards the vulnerability location ($L$), ensuring a targeted and efficient exploration of program behaviors.

*4.2.1 Symbolic Analysis Entrypoint:* To invoke the symbolic execution without any manual intervention, STASE uses the same entry points for symbolic execution used during static analysis. This alignment allows for a seamless transition from static analysis to symbolic execution where $I$ is the field of the vulnerability description $V_{S_P^t}$ is used as the *symbolic analysis entry point*.

*4.2.2 Symbolic Variables:* Symbolic values are allocated to the taint sources identified during static analysis, allowing the symbolic execution engine to explore how external inputs (untrusted data locations) can affect the firmware's execution paths and potentially lead to vulnerabilities. UEFI external inputs typically use a structured data type comprising multiple fields, static analysis reveals the specific fields contributing to the vulnerability. To make the symbolic execution effective, only the contributing fields are defined as symbolic. The $I$ field of the vulnerability description $V_{S_P^t}$ identifies the *symbolic variables* for PEH.

Consider an example involving an SMI handler, SmramProfileHandler, which processes data from a non-SMM environment into an SMM environment and uses CommBuffer as the external input. CommBuffer uses a structure data type comprising multiple fields; however, static analysis reveals that only Command and ReturnStatus, contribute to the targeted vulnerability. Based on this insight, these fields are designated as symbolic within the symbolic execution harness.

```
typedef struct { SMRAM_PROFILE_PARAMETER_HEADER Header;
BOOLEAN RecordingState;
} COMMBUFFER_STRUCT;
COMMBUFFER_STRUCT *CommBuffer=malloc(sizeof(COMMBUFFER_STRUCT));
CommBuffer->Header.Command=klee_int("CommBuffer->Header.Command");
CommBuffer->Header.ReturnStatus=klee_int("CommBuffer->Header.
    ReturnStatus");
```

**Listing 4: Declaring Symbolic Variables**

*4.2.3 Global Variables:* As mentioned earlier, the symbolic execution engine uses the same entry points as static analysis. However, the global variables explored by symbolic execution can be updated in other parts of the code and are accessible from different entry points. This is particularly relevant to cross-handler interactions, where different event handlers and protocols interact through shared global variables. Therefore, relevant global variables must be declared symbolic to allow the symbolic execution engine to reason about the influence of external inputs and internal state changes that occur outside the immediate scope of the currently analyzed code segment.

STASE identifies the relevant global variables as part of its static analysis technique by implementing program slicing to isolate portions of the code relevant to the verification goal. These global variables are then used as the taint sources and included in the $I$ field of the vulnerability description $V_{S_P^t}$. Hence, these global variables are also included in PEH's symbolic variable list.

For example, the global variable mVariableBufferPayload, accessed by multiple handlers and protocols, is identified as a taint source for various entry points. This allows the symbolic execution engine to use it as symbolic and consider various possible states of mVariableBufferPayload, thereby simulating the effects of external inputs and internal state changes across different parts of the code.

*4.2.4 Call Depth:* STASE optimizes the call depth of symbolic execution by managing the breadth of the firmware's codebase, allowing for a more targeted analysis. This optimization minimizes exploring the functions that do not impact the vulnerability under investigation. This process is implemented by taking the output of slicing (the $U$ field of vulnerability description $V_{S_P^t}$) and replacing the functions with stubs. These stubs mimic the basic interface and effects of the original functions but exclude the program logic that is non-critical to the specific vulnerabilities being explored. Consider the following example within the context of SMRAM profile management in the EDKII codebase:

```
1  ContextData = GetSmramProfileContext();
2  if (ContextData == NULL) { return;}
3  SmramProfileGettingStatus = mSmramProfileGettingStatus;
4  mSmramProfileGettingStatus = TRUE;
5  CopyMem(&SmramProfileGetData,SmramProfileParameterGetData,sizeof(
        SmramProfileGetData));
```

**Listing 5: Code snippet containing Non-critical Function**

In this example, GetSmramProfileContext() is responsible for obtaining context data critical for SMRAM profile management. However, for the purposes of vulnerability analysis related to CopyMem, the detailed operations within GetSmramProfileContext() are not directly relevant.

```
MEMORY_PROFILE_CONTEXT_DATA* EFIAPI GetSmramProfileContext(VOID) {
// Initialize a symbolic variable for the context data
MEMORY_PROFILE_CONTEXT_DATA *symbolicContextData;
klee_make_symbolic(&symbolicContextData, sizeof(
    symbolicContextData), "symbolicContextData");
return symbolicContextData;
}
```

**Listing 6: Stub Implementation**

The stub implementation replaces the original function, restricting the call depth within the function. Instead of engaging with detailed firmware data structures, this stub returns a symbolic variable, symbolicContextData. As illustrated in the stub, the minimum implementation is crucial because it maintains control flow integrity. If we remove the function call entirely, the subsequent null check (ContextData == NULL) will fail, disrupting the control flow and leading to inaccurate analysis.

*4.2.5 Assertions.* Static analysis marks the locations as vulnerable where tainted data—data derived from untrusted sources can influence the behavior of the firmware, potentially leading to security breaches. Assertions are added before the taint sinks to enforce constraints or expectations about the program's state at those code points. This helps to eliminate the false positives generated by the static analysis. If an assertion fails during symbolic execution, it indicates a possible vulnerability where attacker-controlled input could manipulate the program's state in unexpected and potentially harmful ways. However, if symbolic execution does not detect an assertion violation, we know that the vulnerability does not exist (within the execution depth explored by symbolic execution).

Steps to add Assertions:

(1) Identify the location: Assertion locations are identified as part of the vulnerability description. This information is provided in the $L$ field of the vulnerability description $V_{S_P^t}$.
(2) Formulate assertions: The static analysis phase of STASE outputs an assertion template (as illustrated in the listing 7) with each vulnerability description $V_{S_P^t}$. The $A$ field holds the assertion template.

The assertion template is filled with the taint sinks to generate the assertion. The $A$ field of vulnerability description $V_{S_P^t}$ holds the taint sinks that can be extracted from the assertion template.

```
Assertion Template: assert(%i >= 0 && %i < Sizeof(%j))
Assertion: klee_assert(arrayIndex >= 0 && arrayIndex <
    bitmapSize) ;
```

**Listing 7: Example of Assertion Template & Assertion**

(3) Implement assertions in code: The assertions are integrated within the firmware code being analyzed.

*4.2.6 Loop Bounds.* Symbolic execution cannot handle unbounded loops. During the static analysis phase of STASE, loops that do not have a constant loop bound are identified and during symbolic execution they are limited a fixed number of loop iterations.

## 5 GUIDED SYMBOLIC EXECUTION

Given a vulnerability description $V_{S_P^t} = \langle P, E, I, A, K, L, U \rangle$, the Environment Configuration Harness (ECH) and the Path Exploration Harness (PEH) instrument the program $P$ and generate an instrumented code segment $\Theta$ with the following properties:

- $\Theta$ is derived from the program $P$, containing only the sliced code without line numbers in $U$, and has specific call depth and loop bounds as specified in PEH.
- $\Theta$ includes a designated program entry point $E$ with a set of symbolic arguments $I$.
- An assertion generated using template $A$ is injected within $\Theta$ at the location $L$.

Guided symbolic execution takes $\Theta$ as input and operates as follows:

(1) **Initialization:** Execution begins with an initial program state $S_P^e$ containing the entry point and symbolic variables.
(2) **Targeted Exploration:** As execution progresses, the symbolic execution engine explores paths with the aim of reaching the location $L$. This involves continually updating through the program state $S_P^i = (M, \overline{\Sigma}) \in \{S_P^e, \ldots, S_P^t\}$ by modifying $M$ (memory state) and $\overline{\Sigma}$ (stack of frames) to reflect only those paths that are relevant to reach instruction $K$. The exploration is constrained by the specified call depth and loop bounds.

(3) **Assertion Validation and State Logging:** Upon reaching $K$, the assertion $A$ is evaluated. If the assertion fails, indicating a vulnerability, the current program state $S_P^t$ is logged. This failure captures the memory state $M$ at the point of failure and adjusts $\overline{\Sigma}$ to reflect the path constraints $\pi$ that led to the vulnerability.

Guided symbolic execution outputs the confirmed vulnerabilities and their signatures. A vulnerability signature consists of the precondition, instrumented code segment containing the vulnerability, and the postcondition describing the immediate program state after triggering the vulnerability.

**Precondition:** The precondition, denoted as $\Pi$, is constructed by combining the set of path constraints for the paths that reach the vulnerability location and results in assertion violation such that $\Pi = \pi_1 \vee \pi_2 \vee \ldots \vee \pi_n$ where each $\pi_i$ is a path constraint.

**Instrumented Code Segment:** The instrumented code segment $\Theta$ is included in the vulnerability signature.

**Postcondition:** The postcondition, denoted as $\Omega$, corresponds to the assertion failure at the vulnerability location.

Combining these components, the vulnerability signature for a specific vulnerability can be represented as a Hoare triple:

$$\{\Pi\} \quad \Theta \quad \{\Omega\}$$

As an example consider an out-of-bounds access vulnerability in the `PxeBcHandleDhcp4Offer` function of EDKII with the following properties:

- Entrypoint: PxeBcHandleDhcp4Offer@PxeBcDhcp4.c:1013
- Assertion Location: PxeBcParseVendorOptions@PxeBcDhcp4.c:232
- Assertion Condition: arrayIndex >= 0 ∧ arrayIndex < bitmapSize

After running a symbolic execution engine on $\Theta$, an assertion failure occurs, indicating the vulnerability. The engine logs three path constraints leading to the assertion condition, which is reported as the precondition ($\Pi$). It also logs the stack traces that reported as the postcondition($\Omega$):

```
1)Precondition:-
(Private->SelectIndex > 0 and Private -> SelectIndex - 1 <
    PXEBC_OFFER_MAX_NUM and and Status = 16)
or (Private -> SelectIndex > 0 and Private -> SelectIndex - 1 <
    PXEBC_OFFER_MAX_NUM and Status = 0 and Private ->
    IsProxyRecved != 0 and Private -> IsOfferSorted != 0 and
    Private -> SelectProxyType < PxeOfferTypeMax and Private ->
    DhcpAck . Dhcp4 -> OptList != 0)
or (Private->SelectIndex > 0  and Private -> SelectIndex - 1 <
    PXEBC_OFFER_MAX_NUM and Status = 14 and Private ->
    IsProxyRecved = 0  and Private -> DhcpAck . Dhcp4 -> OptList
    != 0)
2)Code Segment:-
    Entrypoint: PxeBcHandleDhcp4Offer@PxeBcDhcp4.c:1013
    Symbolic Argument: *Private
    Assertion Location: PxeBcParseVendorOptions@PxeBcDhcp4.c:232
3)Postcondition:-
    !(arrayIndex >= 0 and arrayIndex < bitmapSize) at the program
        location PxeBcParseVendorOptions@PxeBcDhcp4.c:232
```

**Listing 8: Vulnerability Signature**

## 6 EXPERIMENTAL EVALUATION

STASE targets UEFI source code written in C. We use clang-14 to compile the UEFI source code into LLVM-IR. As mentioned above, STASE implementation uses cclyzerpp [32] and Soufflé [33] for rule-based static analysis. We write our vulnerability rules on top of cclyzerpp's analysis rules. We use scripts to generate the environment configuration harness. We automatically generate a driver (path exploration harness) for each entry point based on the vulnerability descriptions computed during the static analysis phase. For symbolic execution, we use KLEE [13] and Z3 [27] for constraint solving. To construct the vulnerability signatures, we use KLEE-generated smt2 [25] and KQuery [10] expressions.

We ran our experiments on a machine with 13th Gen Intel Core i9-13900K CPU at 3.00GHz and 192 GB of RAM running Ubuntu 22.04.4 LTS. STASE implementation is publicly available on GitHub[1]. We experimented on five sets of UEFI implementations to evaluate STASE for UEFI vulnerability detection and characterization. The first and second sets of programs are from the EDKII [44], released as UEFI's "Foundation Code" under an open-source license. These two sets contain the EDKII (release: edk2-stable202311) modules that handle external input:

**EDKII SMM Drivers:** SMM drivers can access highly privileged data and control low-level hardware. SMI Handlers are the most important components of SMM drivers as they are the only channel to receive data from kernel-space programs. This makes them a potential target for attackers who can exploit vulnerabilities in the handlers.

**EDKII Network Module:** The network module processes data received over the network, and it is susceptible to attacks where an attacker can send malicious network packets to exploit vulnerabilities in the network stack. The third and fourth sets of programs (**HARDEN Demo 1**, **HARDEN Demo 2**) were collected from the challenges provided by the Defense Advanced Research Projects Agency (DARPA) for its Hardening Development Toolchains against Emergent Execution Engines (HARDEN) program. These challenge sets use the EDKII source (based on commit af8859bc) and add their own modules and handlers to inject vulnerabilities. In the final set of programs (**Injected EDKII**), we injected five types of vulnerabilities (Use After Free, SMRAM Read and Write, Integer Underflow, Buffer Overflow, and Division by Zero) in the EDKII. We ran our experiments on these five sets of programs using STASE and the following techniques (to compare with STASE's performance):

- **RbSA:** Rule-based Static Analysis with vulnerability rules
- **RbSA-AcTT:** Rule-based Static Analysis with Attacker-controlled Taint Tracking
- **SE:** Symbolic execution using KLEE
- **SE-ECH:** Symbolic execution using KLEE with Environment Configuration Harness
- **HBFA:** Intel's Host-Based Firmware Analyzer [40]

For symbolic execution and HBFA, we limited the experiment to an hour for each entry point. We experimentally evaluate STASE based on the following research questions:

- **RQ1:** To what extent does STASE decrease the false positive rate compared to the rule-based static analysis?
- **RQ2:** Does STASE significantly reduce the time required to identify vulnerabilities compared to traditional methods?
- **RQ3:** How effectively can STASE detect vulnerabilities compared to symbolic execution and fuzzing techniques?

[1]STASE implementation: https://anonymous.4open.science/r/stase-622A.

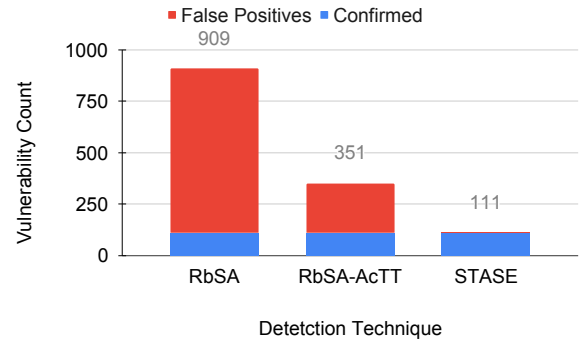

Figure 2: False Positive Reduction by STASE

| Dataset | #EP | SE | | SE-ECH | | HBFA | | STASE | |
|---|---|---|---|---|---|---|---|---|---|
| | | #Vul | Time | #Vul | Time | #Vul | Time | #Vul | Time |
| EDK II SMM | 21 | 0 | 3600 | 0 | 3600 | 1 | 3600 | 13 | 24 |
| EDK II Network | 5 | 0 | 3600 | 0 | 3600 | 0 | 3600 | 5 | 47 |
| HARDEN Demo1 | 27 | 0 | 3600 | 0 | 3600 | 1 | 3600 | 30 | 20 |
| HARDEN Demo2 | 32 | 0 | 3600 | 2 | 3600 | 0 | 3600 | 18 | 28 |
| Injected EDK II | 21 | 0 | 3600 | 2 | 3600 | 1 | 3600 | 45 | 27 |
| Total | 106 | 0 | | 4 | | 3 | | 111 | |

Table 1: Comparison of Vulnerability Detection and Average Analysis Time Per Entrypoint (EP) in Seconds

- **RQ4:** Can STASE precisely identify vulnerability signatures compared to traditional methods?

**RQ1: False positivity reduction by STASE:** Figure 2 shows that STASE eliminates the false positives reported by rule-based static analysis, enhancing the precision of vulnerability detection. We can also see that the false positivity rate is reduced significantly when the vulnerability rules are constrained with attacker-controlled taint tracking without reducing the detection count for true positives. However, this is not enough to achieve full precision, which is only achieved after guided symbolic execution. We also tested Vandalir tool [41] on these datasets. The only matching category of Vandalir with our tool was buffer overflow. It was able to identify 152 targets with none of them being true positive. This is an expected result as Vandalir is not tailored for UEFI.

Table 2 demonstrates that STASE effectively eliminates false positives across all examined vulnerability categories, consistently maintaining a 0% false positivity rate. While rule-based static analysis also achieves 0% in very noticeable patterns such as SMM Callout (triggering EFI_BOOT_SERVICES inside code running in SMM), STASE's superior performance becomes evident in more generic categories like SMRAM READ and WRITE, where it reduces false positives by over 50%. Furthermore, STASE significantly enhances precision in detecting general security vulnerabilities such as integer overflow or buffer overflow, where the static analysis often yields numerous false positives.

**RQ2: Vulnerability detection time reduction by STASE:** Figure 3 compares the average analysis time across different techniques, indicating that STASE finishes exploration on average 27 seconds per entry point which is significantly faster than SE, SE-ECH, and HBFA (all timed out in all cases). Although static analysis methods are faster than STASE, they produce many false positives, as discussed earlier and shown in Figure 2.

| Vulnerability | RbSA | | RbSA-AcTT | | STASE | |
|---|---|---|---|---|---|---|
| | # Vul. | FP % | # Vul. | FP % | # Vul. | FP % |
| SMRAM READ | 261 | 92.34% | 115 | 82.61% | 20 | 0.00% |
| SMRAM WRITE | 501 | 88.02% | 145 | 58.62% | 60 | 0.00% |
| SMM Callout | 2 | 0.00% | 2 | 0.00% | 2 | 0.00% |
| Buffer Overflow | 55 | 70.91% | 33 | 51.52% | 16 | 0.00% |
| Integer Overflow | 14 | 85.71% | 9 | 77.78% | 2 | 0.00% |
| Integer Underflow | 44 | 84.09% | 34 | 79.41% | 7 | 0.00% |
| Out-of-bound Access | 33 | 93.94% | 10 | 80.00% | 2 | 0.00% |
| Use After Free | 2 | 50.00% | 1 | 0.00% | 1 | 0.00% |
| Division by Zero | 20 | 95.00% | 2 | 50.00% | 1 | 0.00% |

**Table 2: Comparison of False Positives (FP) among RbSA, RbSA-AcTT, and STASE**

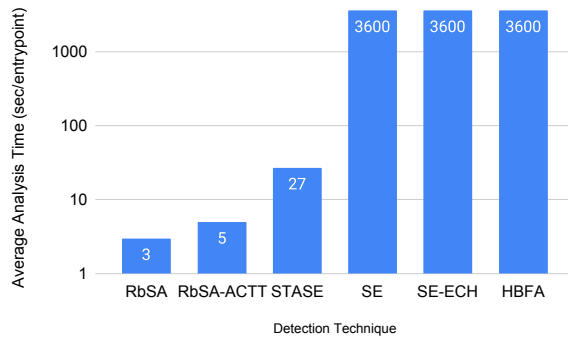

**Figure 3: Comparison of Analysis Time across the Techniques**

**RQ3: Vulnerability detection improvement by STASE:** Table 1 highlights the vulnerability detection improvement by STASE, revealing that STASE detected 111 vulnerabilities across the five datasets, whereas SE, SE-ECH, and HBFA detected far fewer or none. These results highlight that STASE strikes a good balance between reducing the overall analysis time and increasing the number of vulnerabilities detected compared to traditional symbolic execution and fuzzing methods, outperforming both in UEFI vulnerability analysis.

**RQ4: Vulnerability signature generation by STASE:** STASE stands out in its ability to identify vulnerability signatures without requiring any additional analysis time. In our evaluation, STASE successfully identified vulnerability signatures for all 111 vulnerabilities. In contrast, KLEE with ECH was able to generate vulnerability signatures for only 4 vulnerabilities, while vanilla KLEE failed to generate any. Fuzzing methods are not able to generate signatures, as fuzzing typically provides a single input and lacks the capability to generate the precondition and postcondition needed for signature generation. Static analysis methods are not able to generate vulnerability signatures either due to their inherent limitations in capturing the preconditions. This demonstrates the superior precision and effectiveness of STASE in detecting and characterizing vulnerabilities compared to other approaches.

**Discovered Vulnerabilities:** For HARDEN Demo1, HARDEN Demo2, and Injected EDKII, STASE successfully identifies all the injected vulnerabilities. For EDKII SMM drivers, STASE identifies 13 new bugs that we reported to Tianocore. The 5 identified bugs in the EDKII network module correspond to the Pixiefail vulnerabilities.

**Threats to validity:** Our experimental results may be affected by hardware checks, which STASE assumes are sidestepped with specified attacker-controlled inputs. STASE's effectiveness relies on the accuracy of vulnerability rules; if they do not cover a variation, STASE will not identify it. Accurate specification of attacker-controlled entry points and inputs is crucial; otherwise, STASE will identify non-exploitable vulnerabilities. Additionally, the symbolic execution depth and loop bounds we set may also limit the exploration of potential vulnerabilities, as deeper or unbounded loops might contain hidden issues that remain undetected.

## 7 RELATED WORK

UEFI firmware vulnerabilities, though critical, receive less attention than OS and software vulnerabilities. Recent approaches to UEFI vulnerability detection include:

*Symbolic execution.* Bazhaniuk et al. [19] used symbolic execution to find vulnerabilities in UEFI firmware by analyzing SMRAM snapshots, generating 4000 test cases in 4 hours. However, no real bugs were reported, and the manual creation of test harnesses posed a significant challenge. State explosion and high computational costs remain obstacles.

*Fuzzing.* HBFA [40] from Intel has been developed to test UEFI drivers. It supports fuzzing frameworks like AFL, Libfuzzer, and Peach. HBFA's effectiveness is limited by the complexity of adding harnesses for each driver. We compared our approach with HBFA using AFL++ as the underlying fuzzing engine.

*Hybrid analysis.* A hybrid approach called RSFUZZER [49] has been proposed recently, combining fuzzing and symbolic execution; however, it does not produce the vulnerability signatures required for identifying exploitable vulnerabilities. This tool's emulation overhead results in slow performance, achieving only 18-42 executions per second compared to thousands by native fuzzers like AFL. RSFUZZER's effectiveness is also constrained by the availability and completeness of UEFI firmware images, as update bundles often miss important modules. Additionally, it struggles to detect silent vulnerabilities, which do not cause immediate system crashes. Although RSFUZZER claimed their tool would be public, they later indicated it could not be made public, preventing direct comparison.

*Combining static analysis with symbolic execution.* Symbiotic [23] combines static analysis with symbolic execution focusing on property verification and finding a witness that violates a property. STASE is tailored towards generating a vulnerability signature that can be used for characterization. Sys [20] merged user-extensible static checkers with under-constrained symbolic execution to achieve scalability and precision. Their static checkers rely on domain-specific knowledge to detect errors while reducing false positives. False positives still exist after under-constrained symbolic execution. Our technique sidesteps this issue with static and environment configuration harnesses. Recent work [22] on combining static analysis error traces with dynamic symbolic execution demonstrated a negative result. They concluded that the error traces do not provide enough guidance to the path exploration of DSE, and more often than not, static analysis techniques are imprecise. STASE shows improvement in both conclusions. Firstly, we show that combining rule-based static analysis and program slicing is enough to guide SE in achieving scalability. Moreover, we show that the accuracy of the static analyzers can be improved drastically by incorporating the notion of attacker-controlled taint-tracking.

## 8 CONCLUSION

STASE overcomes the challenges of detecting and characterizing UEFI vulnerabilities by combining the strengths of rule-based static analysis and guided symbolic execution. The static analysis phase identifies potential vulnerabilities and outputs detailed descriptions, including entry points, attacker-controlled inputs, and affected instructions. These descriptions are used to automatically generate symbolic execution harnesses, ensuring scalability and eliminating manual intervention. Guided symbolic execution then refines vulnerability detection, reduces false positives, and generates precise vulnerability signatures. STASE demonstrates its effectiveness by successfully identifying injected bugs, 13 new vulnerabilities, and 5 recently reported PixieFail vulnerabilities.

## REFERENCES

[1] Cia vault 7 data leak: What do we know now? https://www.infosecinstitute.com/resources/hacking/cia-vault-7-data-leak-know-since-now/.
[2] Conti leaks reveal ransomware gang's interest in firmware-based attacks. https://thehackernews.com/2022/06/conti-leaks-reveal-ransomware-gangs.html.
[3] Cosmicstrand: the discovery of a sophisticated uefi firmware rootkit. https://securelist.com/cosmicstrand-uefi-firmware-rootkit/106973/.
[4] A deeper uefi dive into moonbounce. https://www.binarly.io/blog/a-deeper-uefi-dive-into-moonbounce.
[5] Finspy: the ultimate spying tool. https://usa.kaspersky.com/blog/finspy-for-windows-macos-linux/25559/.
[6] Hacking team spyware preloaded with uefi bios bootkit to hide itself. https://thehackernews.com/2015/07/hacking-uefi-bios-rootkit.html.
[7] Lojax uefi rootkit overview. https://h20195.www2.hp.com/v2/GetDocument.aspx?docname=4AA7-4019ENW.
[8] Malware delivery through uefi bootkit with mosaicregressor. https://usa.kaspersky.com/blog/mosaicregressor-uefi-malware/23419/.
[9] Mebromi bios rootkit. https://digital.nhs.uk/cyber-alerts/2018/cc-2565.
[10] The reference manual for the kquery language. https://klee-se.org/docs/kquery/.
[11] Trickbot malware gets uefi/bios bootkit feature to remain undetected. https://thehackernews.com/2020/12/trickbot-malware-gets-uefibios-bootkit.html.
[12] Uefi threats moving to the esp: Introducing especter bootkit. https://www.welivesecurity.com/2021/10/05/uefi-threats-moving-esp-introducing-especter-bootkit/.
[13] Klee 3.0, 2023.
[14] L. S. Amour. *Interactive Synthesis of Code-Level Security Rules.* PhD thesis, Northeastern University, 2017.
[15] M. Aung, S. Horwitz, R. Joiner, and T. Reps. Specialization slicing. *ACM Transactions on Programming Languages and Systems (TOPLAS)*, 36(2):1–67, 2014.
[16] G. Balatsouras and Y. Smaragdakis. Structure-sensitive points-to analysis for c and c++. In *Static Analysis: 23rd International Symposium, SAS 2016, Edinburgh, UK, September 8-10, 2016, Proceedings 23*, pages 84–104. Springer, 2016.
[17] R. Barrett and S. Moore. cclyzer++: Scalable and Precise Pointer Analysis for LLVM. https://galois.com/blog/2022/08/cclyzer-scalable-and-precise-pointer-analysis-for-llvm/, 2022.
[18] V. Bashun, A. Sergeev, V. Minchenkov, and A. Yakovlev. Too young to be secure: Analysis of uefi threats and vulnerabilities. In *14th Conference of Open Innovation Association FRUCT*, pages 16–24. IEEE, 2013.
[19] O. Bazhaniuk, J. Loucaides, L. Rosenbaum, M. R. Tuttle, and V. Zimmer. Symbolic execution for BIOS security. In *WOOT*. USENIX Association, 2015.
[20] F. Brown, D. Stefan, and D. Engler. Sys: A {Static/Symbolic} tool for finding good bugs in good (browser) code. In *29th USENIX Security Symposium (USENIX Security 20)*, pages 199–216, 2020.
[21] F. Brown, D. Stefan, and D. R. Engler. Sys: A static/symbolic tool for finding good bugs in good (browser) code. In *USENIX Security Symposium*, pages 199–216. USENIX Association, 2020.
[22] F. Busse, P. Gharat, C. Cadar, and A. F. Donaldson. Combining static analysis error traces with dynamic symbolic execution (experience paper). In *Proceedings of the 31st ACM SIGSOFT International Symposium on Software Testing and Analysis*, pages 568–579, 2022.
[23] M. Chalupa, V. Mihalkovič, A. Řechtáčková, L. Zaoral, and J. Strejček. Symbiotic 9: String analysis and backward symbolic execution with loop folding: (competition contribution). In *International Conference on Tools and Algorithms for the Construction and Analysis of Systems*, pages 462–467. Springer, 2022.
[24] J. Christensen, I. M. Anghel, R. Taglang, M. Chiroiu, and R. Sion. {DECAF}: Automatic, adaptive de-bloating and hardening of {COTS} firmware. In *29th USENIX Security Symposium (USENIX Security 20)*, pages 1713–1730, 2020.
[25] D. R. Cok et al. The smt-libv2 language and tools: A tutorial. *Language c*, pages 2010–2011, 2011.
[26] R. Cytron, J. Ferrante, B. K. Rosen, M. N. Wegman, and F. K. Zadeck. An efficient method of computing static single assignment form. In *Proceedings of the 16th ACM SIGPLAN-SIGACT symposium on Principles of programming languages*, pages 25–35, 1989.
[27] L. De Moura and N. Bjørner. Z3: An efficient smt solver. In *International conference on Tools and Algorithms for the Construction and Analysis of Systems*, pages 337–340. Springer, 2008.
[28] J. Engblom. Finding bios vulnerabilities with symbolic execution and virtual platforms. Last-updated: 06/07/2019.
[29] U. Forum. Uefi specifications. https://uefi.org/specifications.
[30] B. Garmany, M. Stoffel, R. Gawlik, and T. Holz. Static detection of uninitialized stack variables in binary code. In *Computer Security–ESORICS 2019: 24th European Symposium on Research in Computer Security, Luxembourg, September 23–27, 2019, Proceedings, Part II 24*, pages 68–87. Springer, 2019.
[31] S. Horwitz, T. Reps, and D. Binkley. Interprocedural slicing using dependence graphs. *ACM Transactions on Programming Languages and Systems (TOPLAS)*, 12(1):26–60, 1990.
[32] G. Inc. cclyzer++.
[33] H. Jordan, B. Scholz, and P. Subotić. Soufflé: On synthesis of program analyzers. In *Computer Aided Verification: 28th International Conference, CAV 2016, Toronto, ON, Canada, July 17-23, 2016, Proceedings, Part II 28*, pages 422–430. Springer, 2016.
[34] X. Kovah and C. Kallenberg. Are you giving firmware attackers a free pass. In *Proceedings of the RSA Conference, San Francisco, CA, USA*, pages 20–24, 2015.
[35] ldpreload Yukari. BlackLotus UEFI Windows Bootkit. https://github.com/ldpreload/BlackLotus/, 2022.
[36] B. Mullen. Vulnerability management in uefi. https://uefi.org/sites/default/files/resources/Vulnerability%20Management%20in%20UEFI_Mullen.pdf.
[37] D. Oleksiuk. Exploiting ami aptio firmware on example of intel nuc, 2016.
[38] D. Oleksiuk. Thinkpwn. https://github.com/Cr4sh/ThinkPwn/, 2022.
[39] Quarkslab. Pixiefail: Nine vulnerabilities in tianocore's edk ii ipv6 network stack. https://blog.quarkslab.com/pixiefail-nine-vulnerabilities-in-tianocores-edk-ii-ipv6-network-stack.html.
[40] B. Richardson, C. Wu, J. Yao, and V. J. Zimmer. Using host-based firware analysis to improve platform resiliency. Technical report, Intel, 2019.
[41] J. Schilling and T. Müller. Vandalir: Vulnerability analyses based on datalog and llvm-ir. In *International Conference on Detection of Intrusions and Malware, and Vulnerability Assessment*, pages 96–115. Springer, 2022.
[42] Souffle-Lang. Github - souffle-lang/souffle: Soufflé is a variant of datalog for tool designers crafting analyses in horn clauses.
[43] Statista. Global shipments of personal computers. https://www.statista.com/statistics/273495/global-shipments-of-personal-computers-since-2006/.
[44] tianocore. Edk ii. https://github.com/tianocore/tianocore.github.io/wiki/EDK-II.
[45] P. Tsankov. Security analysis of smart contracts in datalog. In *Leveraging Applications of Formal Methods, Verification and Validation. Industrial Practice: 8th International Symposium, ISoLA 2018, Limassol, Cyprus, November 5-9, 2018, Proceedings, Part IV 8*, pages 316–322. Springer, 2018.
[46] M. Weiser. Program slicing. *IEEE Transactions on software engineering*, (4):352–357, 1984.
[47] R. Wojtczuk and C. Kallenberg. Attacks on uefi security. In *Proc. 15th Annu. CanSecWest Conf.(CanSecWest)*, 2015.
[48] Z. Yang, Y. Viktorov, J. Yang, and V. Zimmer. Uefi firmware fuzzing with simics virtual platform. pages 1–6, 07 2020.
[49] J. Yin, M. Li, Y. Li, Y. Yu, B. Lin, Y. Zou, Y. Liu, W. Huo, and J. Xue. Rsfuzzer: Discovering deep smi handler vulnerabilities in uefi firmware with hybrid fuzzing. In *2023 IEEE Symposium on Security and Privacy (SP)*, pages 2155–2169. IEEE, 2023.
[50] J. Yin, M. Li, W. Wu, D. Sun, J. Zhou, W. Huo, and J. Xue. Finding smm privilege-escalation vulnerabilities in uefi firmware with protocol-centric static analysis. In *2022 IEEE Symposium on Security and Privacy (SP)*, pages 1623–1637. IEEE, 2022.
[51] J. Zhao, S. Nagarakatte, M. M. Martin, and S. Zdancewic. Formalizing the llvm intermediate representation for verified program transformations. In *Proceedings of the 39th annual ACM SIGPLAN-SIGACT symposium on Principles of programming languages*, pages 427–440, 2012.